\documentclass[aps,pre,twocolumn,groupedaddress]{revtex4}
\usepackage{graphicx}

\newcommand \be{\begin{eqnarray}}
\newcommand \ee{\end{eqnarray}}

\begin{document}
\title{Stability of negative ionization fronts: 
regularization by electric screening?}
\author{Manuel Array\'as$^{1,2}$ and Ute Ebert$^{2,3}$ }
\affiliation{$^1$Universidad Rey Juan Carlos, Dept.\ de F\'{\i}sica, 
Tulip\'an s/n, 28933, M\'ostoles, Madrid, Spain,}
\affiliation{$^2$CWI, P.O.Box 94079, 1090 GB Amsterdam, The Netherlands,}
\affiliation{$^3$Dept.\ Physics, Eindhoven Univ.\ Techn., The Netherlands.}

\date{\today, accepted for Phys. Rev. E}

\begin{abstract}
We recently have proposed that a reduced interfacial model for streamer 
propagation is able to explain spontaneous branching. Such models require 
regularization. In the present paper we investigate how transversal
Fourier modes of a planar ionization front are regularized by the 
electric screening length. For a fixed value of the electric field 
ahead of the front we calculate the dispersion relation numerically.
These results guide the derivation of analytical asymptotes
for arbitrary fields: for small wave-vector $k$, the growth rate $s(k)$
grows linearly with $k$, for large $k$, it saturates at some
positive plateau value. We give a physical interpretation of these results.
\end{abstract}

\maketitle

\section{Introduction}
Streamers generically appear in electric breakdown
when a sufficiently high voltage is suddenly applied 
to a medium with low or vanishing conductivity.
They consist of extending fingers of ionized matter and
are ubiquitous in nature and technology. Frequently they are 
observed to branch \cite{Williams02,Eddie02}.
There is a traditional qualitative concept for streamer 
branching based on rare photo-ionization events 
\cite{DBM,DBM2,DBM3,Raether,Meek}.
However, our recent work \cite{ME,Andrea,Bernard} has shown 
that even the simplest, fully deterministic streamer model 
without photo-ionization can exhibit branching.
In particular, we have proposed \cite{ME} that a streamer approaching
the Lozansky-Firsov limit of ideal conductivity \cite{Firsov}
can branch spontaneously due to a Laplacian interfacial
instability \cite{Ute}. This mechanism is quite different from 
the one proposed previously. It requires less microscopic physical 
interaction mechanisms, but is based on a dynamically evolving internal 
interfacial structure of the propagating streamer head. Analytical 
branching predictions from the simplest type of interfacial approximation
can be found in \cite{Bernard}.

However, the simple interfacial model investigated in \cite{Bernard}
requires regularization to prevent the formation of cusps. 
The nature of this regularization has to be derived from the 
underlying gas discharge physics; it recently has been subject 
of debate \cite{KuliPRL,ReplyKuli}. We argue that one regularization
mechanism is generically inherent in any discharge model, namely
the thickness of the electric screening layer. This is the subject
of the present paper: we study how the electric screening layer 
present in the partial differential equations of the electric
discharge influences the stability of an ionization front,
correcting the simple interfacial model proposed in
\cite{Firsov,Ute,me,ME} and solved in \cite{Bernard}. 
To be precise, we derive the dispersion relation for transversal
Fourier-modes of a planar ionization front. We treat a negative
front in a model as in \cite{DW,Vit,Ute,me,ME,Andrea}, but with
vanishing electron diffusion and under the assumption
that the state ahead of the ionization front is completely
non-ionized. We have shown previously that the analysis 
of the full model \cite{DW,Vit,Ute,me,ME,Andrea} is mathematically 
nonstandard and challenging due to the ''pulled'' nature 
\cite{pulled1,pulled2} of the front. Pulling is a mode of
front propagation where the spatially half-infinite leading edge 
of a front dominates its behavior. However, for vanishing
electron diffusion and propagation into a non-ionized state, 
the leading edge of the ionization front is completely eliminated and
replaced by a discontinuous jump of the electron density to
some finite value. This corresponds to the fact that neglecting
electron diffusion changes the equation of electron motion from
parabolic to hyperbolic type. Putting $D_e=0$ in the present
paper and considering propagation into a non-ionized state, 
we get rid of leading edge and pulling, but 
in turn we have to analyze discontinuous fronts.

Here we anticipate the result of the paper: if the field far ahead 
of a planar negative ionization front is $E_\infty$, then a transversal 
Fourier perturbation with wave vector $k$ grows with rate
\be
\label{result}
s(k)=\left\{\begin{array}{ll}|E_\infty|k~~&\mbox{for }k\ll\alpha(E_\infty)/2\\
|E_\infty|\alpha(E_\infty)/2~~&\mbox{for }k\gg\alpha(E_\infty)/2\end{array}
\right. ,
\ee
where $\alpha(E)$ is the effective impact ionization coefficient
within a local field $E$; $\alpha$ sets the size of the inverse electric 
screening length. The behavior for large $k$ is a correction to the 
interfacial model treated in \cite{Bernard}; in that model we would have 
$s(k)=|E_\infty|k$ for all $k$. The asymptotes (\ref{result}) 
have been quoted already in \cite{me,ME}, however, without derivation. 
Their derivation based on numerical results and asymptotic analysis
together with a discussion of the
underlying physical mechanisms are the content of the present paper.

In detail, the paper is organized as follows:
in Sec.~II we summarize the minimal streamer model in the limit of 
vanishing diffusion and recall multiplicity, selection and analytical
form of uniformly translating planar front solutions; we then derive
the asymptotic behavior at the position of the shock and far behind
the shock, and we discuss two degeneracies of the problem.
In Sec.~III we set up the framework of the linear perturbation analysis
for transversal Fourier modes, first the equation of motion
and then the boundary conditions and the solution strategy. 
In Sect.~IV we present numerical results for the dispersion 
relation for field $E_\infty=-1$, and we derive the asymptotes 
(\ref{result}) analytically for arbitrary $E_\infty$. The small
$k$ limit is related to one of the degeneracies of the unperturbed 
problem, for the large $k$ limit we also present a physical interpretation. 
Sect.~V contains conclusions and outlook.

\section{Minimal streamer model and planar front solutions}

\subsection{The minimal model}

We investigate the minimal streamer model, i.e., a ``fluid
approximation'' with local field-dependent impact ionization reaction 
in a non-attaching gas like argon or nitrogen
\cite{Rai,DW,Vit,Ute,ME,me,Andrea}.
For physical parameters and dimensional analysis, we refer to
our previous discussions in \cite{Ute,ME,me,Andrea}.
When electron diffusion is neglected ($D_e=0$), 
the dimensionless model has the form
\begin{eqnarray}
\label{107}
\partial_t\;\sigma \;-\; \nabla\cdot\left(\sigma \;{\bf E}\right)
&=& \sigma \; f({\bf E})~,
\\
\label{108}
\partial_t\;\rho 
&=& \sigma \; f({\bf E})~,
\\
\label{109}
\nabla\cdot{\bf E} &=& \rho-\sigma~~,~~{\bf E}=-\nabla \phi~,
\end{eqnarray} 
where $\sigma$ is the electron and $\rho$ the ion density and {\bf E}
the electric field. Here the electron current is assumed to be 
$\sigma {\bf E}$, and the ion current is neglected. Electron--ion pairs 
are assumed to be generated with rate 
$\sigma f({\bf E}) = \sigma|{\bf E}|\;\alpha(|{\bf E}|)$
where $\sigma|{\bf E}|$ is the absolute value of electron current 
and $\alpha({\bf E})$ the effective impact ionization cross section 
within a field ${\bf E}$. Hence $f({\bf E})$ is
\be
\label{f}
f({\bf E})=|{\bf E}|\;\alpha(|{\bf E}|)~.
\ee 
For numerical calculations, we use the Townsend approximation
\begin{equation}
\label{ft}
\alpha(|{\bf E}|)=e^{-1/|{\bf E}|}.
\end{equation}
For analytical calculations, an arbitrary function $\alpha({\bf E})$ 
can be chosen where we only assume that
\be
\label{f2}
f({\bf E})=f(|{\bf E}|)~~\mbox{ and }~~\alpha(0)=0~.
\ee
The last identity entails that $f(0)=0=f'(0)$.
For certain results we also need that $\alpha(|{\bf E}|)$ does not
decrease when $|{\bf E}|$ increases, hence that $\alpha'\ge0$.

Note that the electrons are the only mobile species and the source
of additional ionization, while ion density $\rho$ and 
electric potential $\phi$ or field {\bf E} follow the dynamics 
of the electron density $\sigma$, and couple back onto it.

\subsection{Uniformly translating ionization fronts: 
analytical solutions and multiplicity}
 
We now recall essential properties of uniformly translating planar 
front solutions of Eqs.\ (\ref{107})--(\ref{f}) and (\ref{f2}).
First of all, a constant mode of propagation requires a planar
density distribution that we assume to vary only in the $z$ direction:
$\left(\sigma,\rho\right)=\left(\sigma(z,t),\rho(z,t)\right)$;
the particle densities for large positive $z$ are assumed to vanish.
The field far ahead of the front in the non-ionized region
at $z\to\infty$ has to be constant in time and as a consequence of 
(\ref{109}) also constant in space:
\begin{equation}
\label{1011}
{\bf E}=\left\{\begin{array}{ll} E_\infty\;\hat z~~&~z\to +\infty\\
                                 0~~&~z\to -\infty
               \end{array} \right.~,
\end{equation}
where $\hat z$ is the unit vector in $z$ direction. For the boundary 
condition at $z\to-\infty$ we assumed that the ionized region behind
the front extends to $-\infty$.
This implies, that a fixed amount of charge $\int (\rho-\sigma)dz=E_\infty$
is traveling within the front according to (\ref{109}) and (\ref{1011}), 
and no currents flow far behind the front in the ionized and 
electrically screened region.

For the further analysis, a coordinate system $(x,y,\xi=z-vt)$ 
moving with velocity $v$ in the $z$ direction is used. 
Then the equations (\ref{107})--(\ref{109}) read
\begin{eqnarray}
\label{1010}
\partial_t\sigma -v\partial_\xi\sigma
&-&(\rho-\sigma)\;\sigma +(\nabla\sigma)\cdot(\nabla\phi)
-\sigma f(|\nabla\phi|)=0~,
\nonumber\\
\partial_t\rho -v\partial_\xi \rho &-& \sigma  f(|\nabla\phi|)=0~,
\nonumber\\
&&\rho-\sigma + \nabla^2\phi=0~,
\end{eqnarray} 
where we expressed all quantities by electron density $\sigma$, 
ion density $\rho$ and electric potential $\phi$.

A front propagating uniformly with velocity $v$ is a solution
of (\ref{1011}), (\ref{1010}) where $\sigma$, $\rho$ and $\phi$ 
depend of $\xi$ only. With $\nabla\phi=\partial_\xi\phi\;\hat z=-E \;\hat z$,
such a front solves
\begin{eqnarray}
\label{201}
(v+E)\partial_\xi\sigma+(\rho-\sigma)\;\sigma 
+\sigma f(|E|)&=&0~,
\\
\label{202}
v\partial_\xi \rho + \sigma  f(|E|)&=&0~,
\\
\label{203}
\rho-\sigma-\partial_\xi E&=&0~.
\end{eqnarray} 

For use in the later sections, we now briefly recall the 
analytical solutions \cite{Ute} of these equations.
Subtract (\ref{202}) from (\ref{201}), use (\ref{203})
to get a complete differential, integrate and use
(\ref{203}) again to get $-v\partial_\xi E+\sigma E=\mbox{const.}$.
The integration constant is fixed by the condition $E\to0$ at 
$\xi\to-\infty$ from Eq.~(\ref{1011}), and we find
\begin{equation}
\label{205}
-v\partial_\xi E+\sigma E=0~.
\end{equation}
The front equations then reduce
to two ordinary differential equations for $\sigma$ and $E$
\be
\partial_\xi[(v+E)\sigma]&=&-\sigma f(E)~~,~~f(E)=|E|\alpha(E)~,
\nonumber\\
 v\partial_\xi\ln|E|&=&\sigma,
\ee
that can be solved analytically:
\begin{eqnarray}
\label{2012}
\sigma[E]&=&\frac{v}{v+E}\;\rho[E],\\
\label{2013}
\rho[E]&=&\int^{|E_\infty|}_{|E|}\!\!\frac{f(x)}{x}\;dx
=\int^{|E_\infty|}_{|E|}\!\!\!\!\!\!\alpha(x)dx,\\
\label{2014}
\xi_2-\xi_1&=&\int_{E(\xi_1)}^{E(\xi_2)}\frac{v+x}{\rho[x]}\;\frac{dx}{x}~.\\
\nonumber
\end{eqnarray}
This gives us $\sigma$ and $\rho$ as functions of $E$, 
and the space dependence $E=E(\xi)$ implicitly as $\xi=\xi(E)$ 
in the last equation. It follows immediately from (\ref{2014}) 
that $E(\xi)$ is a monotonic function, and hence that the 
space charge $q=\rho-\sigma=\partial_\xi E$ has the same sign everywhere.
According to (\ref{2013}), $\rho(\xi)$ is a monotonic function, too.

Up to now, the front velocity $v$ as a function of the
asymptotic field $E_\infty$ is not yet fixed.
Indeed for any non-vanishing far field $E_\infty$, there is a continuous
family of uniformly translating front solutions parametrized by $v$
\cite{Ute,Lagarkov},
since the front propagates into an unstable state \cite{pulled1}.
In particular, for $E_\infty>0$ there is a dynamically stable solution 
for any velocity $v\ge0$, and for $E_\infty<0$, there is a dynamically
stable solution for any $v\ge|E_\infty|$. These bounds on
$v$ can be derived directly from Eqs.~(\ref{2012})--(\ref{2014})
with boundary condition (\ref{1011}) and the condition that 
the densities $\sigma$ and $\rho$ are non-negative for all $\xi$.

This continuous family of solutions parametrized by $v$
is associated with an exponentially decaying electron density profile 
in the leading edge \cite{Ute,pulled1}: 
an electron profile that asymptotically for large $\xi$ decays like 
$\sigma(\xi)\propto e^{-\lambda \xi}$ with $\lambda\ge0$,  
will propagate with velocity 
\be
\label{vellam}
v=-E_\infty+\frac{f(E_\infty)}{\lambda} ~\mbox{ in a field }~E_\infty<0.
\ee
It will ``pull'' an ionization front along with the same speed.
(For $E_\infty>0$, the same equation applies for all $\lambda\ge 
f(E_\infty)/E_\infty$, hence for $v\ge0$).

\subsection{Dynamical selection of the shock front solution 
and its particular properties}

In practice, not all these uniformly propagating solutions are observed 
as asymptotic solutions of the full dynamical problem 
(\ref{107})--(\ref{109}), but only a specific one that is called 
the selected front. 
For a negative ionization front, it propagates with the velocity \cite{Ute}
\be
\label{vasym}
v=|E_\infty|~~~\mbox{for }E_\infty<0~.
\ee
The selection takes place through the initial conditions \cite{pulled1}:
If the electron density strictly vanishes beyond a certain point 
$\xi_0$ at time $t=0$
\begin{equation}
\label{1012}\label{init}
\sigma=0=\rho~~\mbox{for }\xi > \xi_0~~\mbox{at }t=0,
\end{equation}
then this stays true for all later times $t$ in the comoving frame $\xi$.
Only initial conditions that decay exponentially like $e^{-\lambda \xi}$
for $\xi\to\infty$, approach a solution with the
larger velocity (\ref{vellam}). Such an exponential decay is a very 
specific initial condition; furthermore, such a leading edge will 
generically be cut off for very small densities by the physical break-down of
the continuum approximation. Therefore the physically relevant solution
is the one with velocity (\ref{vasym}) and absent leading edge 
as in (\ref{init}). The complete absence of the leading edge 
($\lambda=\infty$) is generic for the hyperbolic
equation (\ref{107}), i.e., for vanishing electron diffusion. 
We will restrict the analysis of fronts and their linear 
perturbations to propagation into a completely non-ionized state 
(\ref{init}) in the remainder of the paper.

\begin{figure}[htbp]
  \begin{center}
    \includegraphics[width=0.49\textwidth]{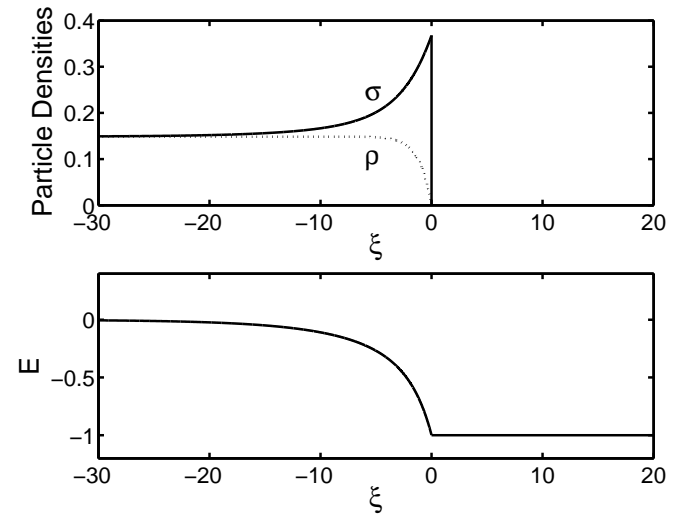} \\
    \caption{Electron density $\sigma$ (solid line in first plot),
  ion density $\rho$ (dotted line in first plot) and electric field 
  $E$ (second plot) for a negative ionization shock front moving with 
  $v=|E_\infty|$ in the comoving frame $\xi=z-vt$. 
  The far field is $E_\infty=-1$.}
   \label{fig1}
  \end{center}
\end{figure}

In contrast to all other uniformly translating fronts with $v>-E_\infty$,
the selected front with $v=-E_\infty$ exhibits a discontinuity 
of the electron density at some point $\xi$ which corresponds to 
$v+E(\xi)\to0$.
We choose the coordinates such that the discontinuity is located at $\xi=0$.
The situation is shown in Fig.~\ref{fig1} for
a uniformly translating front with velocity $v=1$
within a far field $E_\infty=-1$.

A discontinuity of $\sigma$ means that $\partial_\xi\sigma$ 
is singular at this position. On the other hand, the expression
$\sigma(\rho-\sigma+f(E))$ in Eq.\ (\ref{201}) is finite 
or vanishing, therefore the product $(v+E)\partial_\xi\sigma$ 
in Eq.\ (\ref{201}) may not diverge either. Hence $(v+E)$ has 
to vanish at the position of the discontinuity, and therefore 
$E=E_\infty=-v$ at the position of the front.
Furthermore, since $(v+E)\to0$ for $\xi\uparrow0$ while 
$\partial_\xi\sigma$ is bounded for $\xi<0$ ---
as we will derive explicitly below in Eq.\ (\ref{sig-}) 
--- we have 
\be
\label{jap}
\lim_{\xi\to0}\;\;\Big[v+E(\xi)\Big]\;\partial_\xi\sigma=0.
\ee

The fact that $\sigma(\xi)$ in Fig.~\ref{fig1} increases 
monotonically up to the position of the shock, is generic
and can be seen as follows: according to (\ref{201}), 
and since $(v+E)\ge0$ and $\sigma\ge0$, the sign of $\partial_\xi\sigma$
is identical to the sign of $\sigma-\rho-f(E)$. With the help of 
the exact solutions (\ref{2012}) and (\ref{2013}), with the definition
of $f(E)$ in (\ref{f}) and with identifying $v=|E_\infty|$, we find
\begin{equation}
\label{shock}
\sigma-\rho-f(E)=|E|\int_{|E|}^v \frac{\alpha(x)-\alpha(E)}{v-|E|}\;dx\ge0.
\end{equation}
So $\sigma(\xi)$ increases monotonically for growing $\xi$ 
up to $\xi=0$ as long as
$\alpha(E)$ increases monotonically with $E$. This is the case for 
Townsend form (\ref{ft}) or more generally for any $\alpha(E)$
that is monotonically increasing with $E$.

\subsection{Asymptotics near the shock front}

We now derive explicit expressions for $\sigma(\xi)$ etc.\ 
near the discontinuity.
On approaching the position of the ionization shock front
from below $\xi\uparrow0$, the quantity
\be
\label{1eps}
\epsilon=v+E=|E_\infty|-|E|
\ee 
is a small parameter. The ion density (\ref{2013}) at this point 
can be expanded as
\be
\label{1rho}
\rho[E]=\rho[v-\epsilon]=\alpha(v)\epsilon-\alpha'(v)\;\frac{\epsilon^2}{2}
+O(\epsilon^3).
 \ee
As the electron density is related to the ion density through
$\sigma[E]=\rho[E]\;v/\epsilon$ according to (\ref{2012}), it is
\be
\label{1sigma}
\sigma[E]=v\alpha(v)-v\alpha'(v)\frac{\epsilon}{2}+O(\epsilon^2).
\ee
Eq.\ (\ref{2014}) evaluated for $E(\xi_2=0)=E_\infty<0$ reads
\be
-\xi=\int_{|E(\xi)|}^v\frac{v-x}{\rho[x]}\;\frac{dx}{x}
=\int_0^\epsilon \frac{y}{\rho[v-y]}\;\frac{dy}{v-y},
\ee
where in the last expression, the parameter $\epsilon$ (\ref{1eps})
is introduced. Insertion of (\ref{1rho}) now yields an explicit
relation between $\xi$ and $E$:
\be
-\xi&=&\frac{\epsilon}{v\alpha(v)}+O(\epsilon^2)\\
\mbox{or}~~\epsilon&=&-v\alpha(v)\xi+O(\xi^2).
\nonumber
\ee
Insertion of this approximately linear relation between $\epsilon$
and $\xi$ into (\ref{1rho}) and (\ref{1sigma}) together with the notation
$f(v)=v\alpha(v)$ results in
\begin{eqnarray}
\label{sig-}
\sigma(\xi)&=&\theta(-\xi)
\left(f(v)+\frac{f(v)\;v\;\alpha'(v)}{2}\;\xi+O(\xi^2)\right),\\
\label{rho-}
\rho(\xi)&=&\theta(-\xi)\Big(-f(v)\;\alpha(v)\;\xi+O(\xi^2)\Big),\\
\label{E-}
-E(\xi)&=&v+\theta(-\xi)\Big(f(v)\;\xi+O(\xi^2)\Big),
\end{eqnarray}
where we used $v=|E_\infty|$ and the step function $\theta(x)$,
defined as $\theta(x)=1$ or $0$ for $x>0$ or $x<0$, respectively.

\subsection{Asymptotics far behind the shock front}

Far behind the front in the ionized region $\xi\to-\infty$, the fields
approach $\lim_{\xi\to-\infty}(\sigma,\rho,E)=(\sigma^-,\rho^-,E^-)$ with
\begin{equation}
\label{infvalue}
\sigma^-=\rho^-=\int_0^v\alpha(x)\;dx~~,~~E^-=0~.
\end{equation}
Expanding about this point as $\sigma(\xi)=\sigma^-+\sigma_1(\xi)$ etc.,
we derive in linear approximation
\begin{eqnarray}
\label{eq}
\partial_\xi \left(\begin{array}{c}\sigma_1\\ \rho_1\\ -E_1\end{array}\right)=
\left(\begin{array}{ccc}
  \lambda & -\lambda & 0\\ 
  0 &  0 & 0 \\
  1 & -1 & 0
\end{array}\right)
\cdot
\left(\begin{array}{c}\sigma_1\\ \rho_1\\ -E_1\end{array}\right)~,
\end{eqnarray}
with $\lambda$ given by
\be
\label{lambda}
\lambda=\frac{\sigma^-}{v}=
\int_0^v\alpha(x)\;\frac{dx}{v}~.
\ee
Two eigenvalues of the matrix in (\ref{eq}) vanish. The third eigenvalue 
of the matrix is the positive parameter $\lambda$, it produces 
the eigendirection
\be
\label{lim0-}
\left(\begin{array}{c}\sigma\\ \rho\\-E\end{array}\right)(\xi)
&=&\left(\begin{array}{c}\sigma^-\\\sigma^-\\0\end{array}\right)
+A\;\left(\begin{array}{c}\lambda\\0\\1\end{array}\right)
\;e^{\:\lambda \xi}+O\left(e^{\:2\lambda \xi}\right)~,
\nonumber\\
&&\mbox{for }~\xi\to-\infty,
\end{eqnarray}
that describes the asymptotic solution deep in the ionized region.
The free parameter $A>0$ accounts for translation invariance.

\subsection{Two degeneracies of the shock front}

We have fixed the initial condition (\ref{init}) and hence 
we have selected the front speed $v=-E_\infty$. Therefore 
the degeneracy of solutions related to the profile of the
leading edge is removed. Still there
are two degeneracies remaining in the problem. The first
one is the well known mode of infinitesimal translation
that corresponds to the arbitrary position of the front.
The second one is specific for the present problem and will 
play a central role in the derivation of the analytical asymptote
for small $k$ in Sect.~IV.
It is the mode of infinitesimal change of far field $E_\infty$. 
It corresponds to the arbitrariness of the field $E_\infty$
in the non-ionized region with $\sigma=0=\rho$ ahead of the front
and to the corresponding arbitrariness of the asymptotic ionization 
level $\sigma=\sigma^-=\rho$ behind the front where the field vanishes.
To set the stage for the later analysis, the necessary
properties of the modes are given.

An infinitesimal translation of the front in space
generates the linear mode $(\sigma_t,\rho_t,E_t)=
(\partial_\xi\sigma,\partial_\xi\rho,\partial_\xi E)$
\begin{eqnarray}
(v+E)\partial_\xi\sigma_t&=&\Big(2\sigma-\rho-f\Big)\sigma_t
-\sigma\rho_t
+\Big(\sigma f'-\partial_\xi\sigma\Big)E_t
\nonumber\\
v\partial_\xi\rho_t&=&-f\sigma_t+\sigma f'E_t
\nonumber\\
\partial_\xi E_t&=&\rho_t-\sigma_t
\end{eqnarray}
with the definition $f'=\partial_xf(|x|)$, so that 
$f(E+E_t)=f-f'E_t+\ldots$ for $E<0$. With the notation
$\psi_t=-E_t$, the equations can be written in matrix form as
\begin{eqnarray}
\label{eq0}
&&\partial_\xi \left(\begin{array}{c}\sigma_t\\ \rho_t\\ \psi_t
                     \end{array}\right)
={\bf N}_0(\xi)\cdot
\left(\begin{array}{c}\sigma_t\\ \rho_t\\ \psi_t\end{array}\right)
\\
\label{matr03}
&&{\bf N}_0(\xi)=\left(\begin{array}{ccc}
  \displaystyle\frac{2\sigma-f-\rho}{v+E}&\displaystyle\frac{-\sigma}{v+E}&
      \displaystyle\frac{\partial_\xi\sigma-\sigma f'}{v+E}\\ 
  &&\\
  \displaystyle\frac{-f}{v}&\displaystyle0&\displaystyle\frac{-\sigma f'}{v}\\ 
  &&\\
  \displaystyle1&\displaystyle-1&\displaystyle0
\end{array}\right)~.
\end{eqnarray}
Note that the matrix ${\bf N}_0(\xi)$ reduces to the matrix in 
Eq.\ (\ref{eq}) for $\xi\to-\infty$, since $(2\sigma-f-\rho)/(v+E)
\to \sigma^-/v=\lambda$ etc. The limiting value of the vector 
$(\sigma_t,\rho_t,\psi_t)$ for
$\xi\to0$ is according to (\ref{sig-})--(\ref{E-})
\begin{equation}
\left(\begin{array}{c}\sigma_t\\ \rho_t\\ \psi_t\end{array}\right)
\stackrel{\xi\uparrow0}{\longrightarrow}
\left(\begin{array}{c}
fv\alpha'/2\\ -f\alpha \\ f\end{array}\right).
\end{equation}

The second mode is generated by an infinitesimal change of the far
field $E_\infty$ and consecutively by an infinitesimal change
of the velocity $v$. The discontinuity is taken
at the position $\xi=0$. In linear order, this variation creates
a mode 
\be
\label{sigE}
\sigma_E(\xi)=\lim_{\epsilon\to0}
\frac{\sigma^{[E_\infty+\epsilon]}(\xi)-\sigma^{[E_\infty]}(\xi)}{\epsilon}
~~~\mbox{etc.},
\ee
that solves the inhomogeneous equation
\begin{equation}
\label{2033}
\partial_\xi \left(\begin{array}{c}\sigma_E\\ \rho_E\\ \psi_E\end{array}\right)
={\bf N}_0(\xi)\cdot
\left(\begin{array}{c}\sigma_E\\ \rho_E\\ \psi_E\end{array}\right)
- \left(\begin{array}{c}\partial_\xi\sigma/(v+E)\\ \partial_\xi\rho/v\\ 0 
        \end{array}\right)~.
\end{equation}
The inhomogeneity vanishes at $\xi\to-\infty$. Hence like the front
solution itself and like the infinitesimal translation mode, also
this mode has the eigendirection 
$(\delta\sigma^-,\delta\sigma^-,0)+A\;(\lambda,0,1) \;e^{\lambda \xi}+\ldots$ 
asymptotically for $\xi\to-\infty$. The value of $\delta\sigma^-$
is given by $\delta\sigma^-= \partial\sigma^-/\partial |E_\infty|=
\alpha(E_\infty)$ according to (\ref{infvalue}). For $\xi\uparrow0$, 
the limiting values of the fields are
\begin{equation}
\label{bc01}\label{2034}
\left(\begin{array}{c}\sigma_E\\ \rho_E\\ \psi_E\end{array}\right)
\stackrel{\xi\uparrow0}{\longrightarrow}
\left(\begin{array}{c}
f'\\ 0 \\ 1\end{array}\right),
\end{equation}
which is the derivative of (\ref{sig-})--(\ref{E-}) with respect 
to $v=|E_\infty|$ at $\xi=0$.

\section{Set-up of linear stability analysis}

We now can proceed to study the stability of a planar ionization shock front.
The front propagates into the $z$ direction. The perturbations have 
an arbitrary dependence on the transversal coordinates $x$ and $y$.
Within linear perturbation theory, they can be decomposed into
Fourier modes. Therefore we need the growth rate $s(k)$ of an arbitrary 
transversal Fourier mode to predict the evolution of an
arbitrary perturbation. Because of isotropy within the transversal 
$(x,y)$-plane, we can restrict the analysis to Fourier modes in the 
$x$ direction, so we study linear perturbations $\propto e^{st+ikx}$.
(The notation anticipates the exponential temporal growth of such modes.)

In general, there can be a degeneracy of the dispersion relation
$s(k)$ for various profiles of the leading edge just as it is found 
also for the uniformly translating solutions in Section II.B. 
The constraint of a non-ionized initial condition (\ref{init}) 
again will remove this degeneracy and fix $s(k)$.
In the present section, we will derive the equations and the boundary 
conditions for the Fourier modes. In Sect.\ IV, we will solve them 
numerically and derive the analytical asymptotes (\ref{result}). 

\subsection{Equation of motion}

The linear perturbation theory could be set up within 
the coordinate system $(x,\xi=z-vt)$ that moves with 
the unperturbed constant velocity $v=|E_\infty|$.
This would, of course, lead to a set of equations that are
linear in the perturbation. 

However, when the perturbation of a planar front grows,
the position of the actual discontinuity of the electron
density will deviate from the position of the discontinuity
of the unperturbed front. Within the coordinate system
$(x,\xi)$, this would lead to finite deviations within
infinitesimal spatial intervals instead of infinitesimal
deviations within finite intervals. This conceptual
difficulty can be avoided by formulating the perturbation theory 
within the coordinate system of the position of the perturbed 
shock front $(x,\zeta)$ with
\be
\label{zeta}
\zeta=\xi-\Delta_k~~,~~ \xi=z-vt~~,~~\Delta_k=\delta\;e^{ikx+st}~,
\ee
where $z$ is the rest frame, $\xi$ is the frame moving with the
planar front and $\zeta=0$ marks the line of electron discontinuity 
of the actual front. Therefore we write the perturbation as
\begin{eqnarray}
\label{pertu}
\sigma(x,\zeta,t)&=&\sigma_0(\zeta)+\sigma_1(\zeta)\;\Delta_k(x,t)
\nonumber\\
\rho(x,\zeta,t)&=&\rho_0(\zeta)+\rho_1(\zeta)\;\Delta_k(x,t),
\nonumber\\
\phi(x,\zeta,t)&=&\phi_0(\zeta)+\phi_1(\zeta)\;\Delta_k(x,t),
\end{eqnarray}
where $\sigma_0$, $\rho_0$ and $\phi_0$ are the electron density, 
ion density and electric potential of the planar ionization shock
front from the previous section. But these planar
solutions here are shifted to the position of the perturbed front $\zeta$.
Therefore they do not move with their proper velocity $v=-\partial_t\xi$,
but with a slightly different velocity $-\partial_t\zeta=v-s\Delta_k$.
The price to pay is that the equations of the perturbation analysis
become inhomogeneous, actually in a similar way as in \cite{pulled1}.
The gain is that the derivation of the boundary conditions at
the shock front becomes more comprehensible, and that later
in Section V.B the identification of the analytical solution
for small $k$ with the mode $(\sigma_E,\rho_E,\psi_E)$ from the
previous section becomes quite obvious.

Substitution of the expressions (\ref{pertu}) into (\ref{1010}) 
gives to leading order in the small parameter $\delta$
\begin{eqnarray}
\label{EQ}
(v+E_0)\;\partial_{\zeta}\sigma_1&=&(s+2\sigma_0-\rho_0-f)\;\sigma_1
\nonumber\\
& &-\sigma_0\;\rho_1+(\partial_{\zeta}\sigma_0-\sigma_0 f')\;
\partial_{\zeta} \phi_1-s\partial_{\zeta}\sigma_0,
\nonumber\\
v\;\partial_{\zeta}\rho_1&=&-f\;\sigma_1
+s\;\rho_1-\sigma_0 f'\;\partial_{\zeta} \phi_1-s\partial_{\zeta}\rho_0,
\nonumber\\
\left(\partial_{\zeta}^2-k^2\right)\;\phi_1&=&\sigma_1-\rho_1+k^2E_0.
\end{eqnarray}
Here $f=f(E_0)$, $f'=\partial_{|E|}f(|E|)\Big|_{E_0}$, and 
$E_0 = -\partial_{\zeta} \phi_0(\zeta)$ 
is the electric field of the uniformly translating front. As explained above,
these equations are not completely linear in $(\sigma_1,\rho_1,\phi_1)$, 
but contain the inhomogeneities $s\partial_\zeta\sigma_0$, 
$s\partial_\zeta\rho_0$ and $k^2E_0$. 

To elucidate the structure of Eq.\ (\ref{EQ}), we drop all indices 0
and introduce the matrix notation
\begin{eqnarray}
\label{Matr1}
&&\partial_{\zeta} 
\left(\begin{array}{c}\sigma_1\\ \rho_1\\ \psi_1\\ \phi_1\end{array}\right)
={\bf M}_{s,k}\cdot
\left(\begin{array}{c}\sigma_1\\ \rho_1\\ \psi_1\\ \phi_1\end{array}\right)
- \left(\begin{array}{c}s\partial_{\zeta}\sigma/(v+E)\\ 
                        s\partial_{\zeta}\rho/v\\ - Ek^2\\0 
        \end{array}\right),
\nonumber\\
&&\\
\label{Matr2}
&&{\bf M}_{s,k}(\zeta)=\left(\begin{array}{cccc}
  \displaystyle\frac{s+2\sigma-f-\rho}{v+E}
  &\displaystyle\frac{-\sigma}{v+E}&
      \displaystyle\frac{\partial_{\zeta}\sigma-\sigma f'}{v+E}&0\\ 
  &&&\\
  \displaystyle\frac{-f}{v}&\displaystyle\frac{s}{v}&
      \displaystyle\frac{-\sigma f'}{v}& 0\\ 
  &&&\\
  \displaystyle1&\displaystyle-1&\displaystyle0&k^2\\
  &&&\\
  0&0&1&0
\end{array}\right).
\nonumber\\
\end{eqnarray} 
Here we introduced the auxiliary field 
\be
\label{psi1}
\psi_1 = \partial_{\zeta} \phi_1,
\ee
that corresponds to the perturbation $E_1$ of the electric field,
but with reversed sign.

\subsection{Boundary conditions at the discontinuity}

Having obtained the perturbation equations, we are now in the position 
to derive the boundary conditions. First we consider the boundary 
conditions at $\zeta=0$ where we make explicit use of the initial
condition (\ref{init}). The boundary conditions arise from the 
boundedness of the electron density to the left of the shock front
at $\zeta\uparrow0$, and from the continuity of all other fields 
across the position $\zeta=0$ of the shock front.

As discussed in Section II.D, for the uniformly propagating
shock front, the quantity $(v+E)\;\partial_{\zeta}\sigma$
vanishes as $\zeta\uparrow0$, since $(v+E)$ vanishes and
$\partial_\zeta\sigma$ is bounded. Since this should hold
both for the full solution as well as for the unperturbed solution,
it also holds for the perturbation
\be
\label{limsig-}
\lim_{\zeta\uparrow0}\;\;\Big[v+E(\zeta)\Big]\;\partial_\zeta\sigma_1=0.
\ee
Furthermore
\be
\label{limsig+}
\Big[v+E(\zeta)\Big]\;\partial_\zeta\sigma_1\equiv 0~~\mbox{for }\zeta\ge0.
\ee
This identity is trivial for $\zeta>0$, but nontrivial for $\zeta=0$.
When the explicit expressions
(\ref{sig-})--(\ref{E-}) are inserted into (\ref{EQ}), we find
\be
(v+E)\partial_\zeta\sigma_1&=&
\big(s+f(v)\big)\sigma_1-f(v)\rho_1-f(v)f'(v)\psi_1
\nonumber\\
&&+(\psi_1-s)\;\partial_\zeta\sigma+O(\zeta).
\ee
First of all, $\partial_\zeta\sigma$ is singular at $\zeta=0$,
since $\partial_\zeta\sigma\propto \partial_\zeta\theta(-\zeta)
=-\delta(\zeta)$. Therefore (\ref{limsig+}) requires 
that the coefficient of $\partial_\zeta\sigma$ must vanish
\be
\label{bc11}
\psi_1(0)=s
\ee
which gives the first boundary condition. Second, applying now
(\ref{limsig-}) yields the second boundary condition
\be
\label{bc12}
\big(s+f(v)\big)\sigma_1(0)-f(v)\rho_1(0)-f(v)f'(v)\psi_1(0)=0.
\ee
Due to the discontinuity, actually two boundary conditions
(\ref{bc11}) and (\ref{bc12}) result from (\ref{limsig-}) and
(\ref{limsig+}).

In a second step the continuity of the other fields across $\zeta=0$
is evaluated. The continuity of $\rho$ we get from (\ref{202}) 
and the fact, that $\sigma$ and $|{\bf E}|$ are bounded for all $\xi$.
It immediately yields the third boundary condition
\be
\label{bc13}
\rho_1(0)=0,
\ee
just like for the unperturbed equation. Finally, for the 
boundary conditions on field and potential, it is helpful 
that there is an exact solution for the non-ionized region at 
$\zeta>0$ for a boundary with the harmonic form (\ref{zeta}).
Since ahead of the front there are no particles $\sigma=0=\rho$,
there are also no space charges, and for the potential,
one has to solve $\nabla^2\phi=0$ with the limit 
${\bf E}=-\nabla \phi \to E_\infty\;\hat \zeta = - v ~\hat \zeta$ 
as $\zeta\to\infty$. The general solution for $\zeta>0$ is
\be
\label{xig0}
\phi&=&v\xi+\delta\;c\;e^{-k\xi}\;e^{ikx+st}
\nonumber\\
&=&v\zeta+\delta(v+c\;e^{-k\zeta})\;e^{ikx+st}+O(\delta^2)
\ee
with the yet undetermined integration constant $c$.
Here we chose the gauge $\phi_0(\xi=0)=0$ for the unperturbed
electric potential.

Now $\phi$ always is continuous, and ${\bf E}=-\nabla\phi$ 
is continuous, because the charge density $|\rho-\sigma|<\infty$ 
in (\ref{109}) everywhere. The continuity of $\phi$ at $\zeta=0$ implies
\be
\label{bc14}
\phi_1(0)=v+c,
\ee
the continuity of $\partial_x\phi$ yields the same condition,
and the continuity of $\partial_\zeta\phi$ implies
\be
\label{bc15}
\psi_1(0)=-ck.
\ee

The five boundary conditions (\ref{bc11})--(\ref{bc13}) and 
(\ref{bc14})--(\ref{bc15}) determine the value of the 
integration constant 
\be
\label{c}
c=-\;\frac{s}{k}
\ee 
in (\ref{xig0}) and the values of the four fields at $\zeta=0$ 
\begin{equation}
\label{Init}
\left(\begin{array}{c}\sigma_1\\ \rho_1\\ \psi_1\\ \phi_1\end{array}\right)
\stackrel{\zeta\uparrow0}{\longrightarrow}
\left(\begin{array}{c}f'(v)\;sf(v)/(s+f(v))\\ 0\\ s\\ (vk-s)/k
\end{array}\right)~.
\end{equation}
Hence the explicit solution in the non-ionized region $\zeta>0$ is
\be
\label{sol+}
\sigma(x,\zeta>0,t)&=&0=\rho(x,\zeta>0,t),\\
\phi(x,\zeta>0,t)&=&v\zeta+\delta\;\frac{vk-s\;e^{-k\zeta}}{k}\;e^{ikx+st}
+O(\delta^2).
\nonumber
\ee

\subsection{Solution strategy and limits for $\zeta\to-\infty$}

We aim to calculate the dispersion relation $s=s(k)$ for fixed $k$.
For any $s$ and $k$, the solution at $\zeta>0$ is given explicitly
by (\ref{sol+}). This solution determines the value of the fields
(\ref{Init}) at $\zeta=0$ as a unique function of $s$ and $k$. 
The expression (\ref{Init}) is the initial condition for 
the integration of (\ref{Matr1}) towards $\zeta\to-\infty$.
The requirement that the solution approaches a physical limit
at $\zeta\to-\infty$ has to determine $s$ as a function of $k$.
According to a counting argument, this is indeed the case,
as will be explained now. 

First, the limiting values of the fields at $\zeta=-\infty$ are 
comparatively easy:
the total charge vanishes, hence $\sigma_1$ and $\rho_1$ approach the
same limiting value $\sigma_1\to\sigma_1^-$ and $\rho_1\to\sigma_1^-$, 
and the electric field vanishes, hence $\psi_1\to0$ and $\phi_1\to0$. 
Here the limiting values at $\zeta\to-\infty$ again were denoted 
by the upper index ${}^-$ as in (\ref{lim0-}).

Second, the eigendirections are determined by linearizing the
equations of motion (\ref{Matr1}) about this asymptotics.
In a calculation similar to the one from Sect.~II.F, one derives
for $\zeta\to-\infty$
\begin{eqnarray}
\label{End}
\left(\begin{array}{c}\sigma_1\\ \rho_1\\ \psi_1\\ \phi_1\end{array}\right)
\displaystyle\stackrel{\zeta\to-\infty}{\approx}&&
\left(\begin{array}{c}\sigma_1^-\\ \sigma_1^-\\ 0\\ 0\end{array}\right)
\nonumber\\
&&+a_1\;e^{\lambda_1\zeta}
\left(\begin{array}{c}\lambda_1^2-k^2\\ 0\\ \lambda_1\\ 1\end{array}\right)
+ a_2\;e^{\lambda_2\zeta}
\left(\begin{array}{c}1\\ 1\\ 0\\ 0\end{array}\right)
\nonumber\\
&&+ a_3\;e^{k\zeta}
\left(\begin{array}{c}0\\ 0\\ k\\ 1\end{array}\right)
+ a_4\;e^{-k\zeta}
\left(\begin{array}{c}0\\ 0\\ -k\\ 1\end{array}\right)
\end{eqnarray}
with the free parameters $a_1, a_2, a_3, a_4$ and $\sigma_1^-$ 
and the eigenvalues
\begin{equation}
\label{end3}
\lambda_1=\frac{\sigma^-+s}{v}=\lambda+\lambda_2~~,~~\lambda_2=\frac{s}{v}
\end{equation}
and $\lambda$ from Eq.\ (\ref{lambda}).

For positive $s$ and $k$, all eigenvalues $\lambda_1$, $\lambda_2$
and $k$ are positive except for the fourth one $-k$. Hence
the first three eigendirections approach the appropriate limit
for $\zeta\to-\infty$, while the fourth one does not.
Therefore a solution can only be constructed for 
\be
\label{a4}
a_4=0.
\ee
This condition determines the dispersion relation $s=s(k)$ 
when a solution of (\ref{Matr1}) and (\ref{Init}) is integrated
towards $\zeta\to-\infty$.

\section{Calculation of the dispersion relation}

Having set the stage, the dispersion relation is now first 
evaluated numerically for $E_\infty=-1$. Besides an expected
result for small $k$, this investigation has delivered 
a previously unexpected result for large $k$. Based on
these numerical results for fixed $E_\infty$, we were able 
to derive analytical asymptotes for small or large $k$ and 
for arbitrary $E_\infty<0$. We also understood the physical 
mechanism driving this asymptotic behavior. The section contains
the derivation of our numerical results and of our analytical asymptotes 
and their physical interpretation.

\subsection{Numerical results for arbitrary $k$ and $E_\infty=-1$}

The problem is to integrate the equations for the transversal perturbation 
(\ref{Matr1}) for fixed $k$ and guessed $s$ from the initial condition 
(\ref{Init}) at $\zeta=0$ towards decreasing $\zeta$.
In general, the boundary condition (\ref{End}) with (\ref{a4})
will not be met, so $s$ has to be iterated until $a_4\approx0$.
When the condition is met, the solution does not diverge for large 
negative $\zeta$, otherwise it does. When passing through the
appropriate $s=s(k)$, the sign of the divergence changes.
This is how the data of Fig.~2 with their error bars were derived.

For the numerical integration, the ODEPACK collection of subroutines 
for solving initial value problems was used \cite{Hind} to solve
the seven ordinary differential equations for the unperturbed problem
(\ref{201})--(\ref{203}) and the perturbation (\ref{Matr1})--(\ref{Matr2})
 simultaneously. The unperturbed solution has to be calculated
since it enters the matrix (\ref{Matr2}).

However, the numerics can not directly be applied to the problem
in the form (\ref{Matr1})--(\ref{Matr2}) because the matrix contains
apparently diverging terms proportional to $1/(v+E(\zeta))$ for $\zeta\to0$.
Therefore the behavior of the solution for $\zeta\to0$ has to
be evaluated in a similar way as in Sect.~II.E. With the ansatz 
\begin{eqnarray}
\label{exp1}
\sigma_1(\zeta)=\sigma_1(0^-)+C_1 \zeta+O(\zeta^2)~,\nonumber\\
\rho_1(\zeta)=\rho_1(0^-)+C_2 \zeta+O(\zeta^2)~,\nonumber\\
\psi_1(\zeta)=\psi_1(0^-)+C_3 \zeta+O(\zeta^2)~,\nonumber\\
\phi_1(\zeta)=\phi_1(0^-)+C_4 \zeta+O(\zeta^2)~,
\end{eqnarray}  
where $\sigma_1(0^-)$ etc.\ are given by (\ref{Init}),
the parameters $C_i$ become
\begin{eqnarray}
\label{exp2}
C_2=-s\alpha\left(\frac{ff'}{s+f}+f+f'\right)~,\nonumber\\
C_3=s\left(-k+\frac{ff'}{s+f}\right)~~,~~C_4=s~,\\
C_1=\frac{C_2+(\alpha+v\alpha'/2)\;C_3+s(v\alpha f''+v\alpha'f'/2)}{2+s/f}
\nonumber
\end{eqnarray}
In the numerical procedure, the explicit solutions (\ref{sig-})--(\ref{E-})
and (\ref{exp1})--(\ref{exp2}) are used until $\zeta=10^{-5}$, then 
the differential equations are evaluated. 

 \begin{figure}[htbp]
  \begin{center}
    \includegraphics[width=0.49\textwidth]{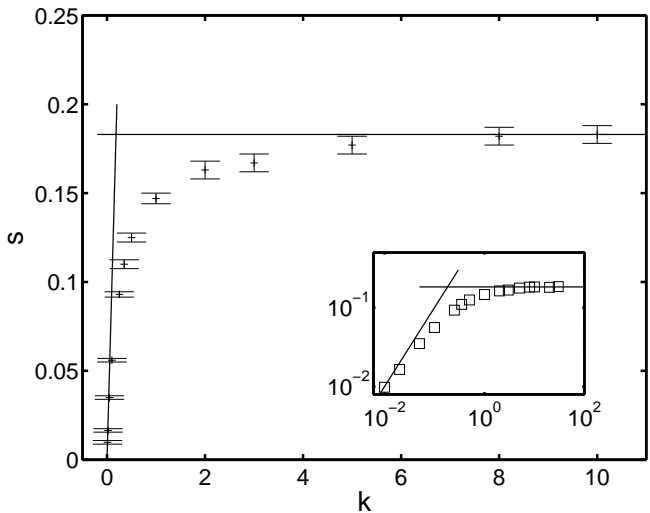} 
    \caption{Dispersion curve for $E_\infty=-1$, hence $v=1$.
    The big figure shows the numerical data with error bars
    and the two analytical asymptotes for small and large $k$ (lines). 
    The inset shows the same data (squares) in double-logarithmic 
    scale with the same two analytical asymptotes.}
   \label{fig2}
  \end{center}
 \end{figure}

The numerical results for the dispersion relation in a field $E_\infty=-1$,
i.e., for a shock front with velocity $v=1$ are shown in Fig.~\ref{fig2}.
It can be seen that the dispersion curve for small $k$ grows linearly, 
but then turns over and finally for large $k$ saturates at a constant value. 

\subsection{Asymptotics for small $k$ and arbitrary $E_\infty<0$}

We first derive the asymptotic behavior for small $k$ 
for an arbitrary far field $E_\infty<0$.

When the equations of motion (\ref{Matr1}) and (\ref{Matr2}) 
are evaluated up to first order in $k$, $\phi_1$ decouples, and we get
\begin{equation}
\label{3031}
\partial_{\zeta} \left(\begin{array}{c}\sigma_1\\ \rho_1\\ \psi_1
                       \end{array}\right)
={\bf N}_s\cdot
\left(\begin{array}{c}\sigma_1\\ \rho_1\\ \psi_1\end{array}\right)
- \left(\begin{array}{c}s\partial_{\zeta}\sigma/(v+E)\\ 
                        s\partial_{\zeta}\rho/v\\ 0 
        \end{array}\right)+O(k^2)~,
\end{equation}
where 
\begin{equation}
\label{3032}
{\bf N}_s(\zeta)
=\left(\begin{array}{ccc}
  \displaystyle\frac{s+2\sigma-f-\rho}{v+E}&\displaystyle\frac{-\sigma}{v+E}&
      \displaystyle\frac{\partial_{\zeta}\sigma-\sigma f'}{v+E}\\ 
  &&\\
  \displaystyle\frac{-f}{v}&\displaystyle\frac{s}{v}&
      \displaystyle\frac{-\sigma f'}{v}\\ 
  &&\\
  \displaystyle1&\displaystyle-1&\displaystyle0\\
\end{array}\right)+O(k^2)
\end{equation}
is the truncated matrix ${\bf M}_{s,k}(\zeta)$ (\ref{Matr2}). The matrix 
${\bf N}_s$ for $s=0$ reduces to the matrix ${\bf N}_0$ from Eq.\
(\ref{matr03}) --- this fact will be instrumental below. 
The fourth decoupled equation reads
\begin{equation}
\label{3033}
\partial_{\zeta}\phi_1=\psi_1
\end{equation}
The boundary condition (\ref{Init}) reduces to
\begin{equation}
\label{3034}
\left(\begin{array}{c}\sigma_1\\ \rho_1\\ \psi_1\end{array}\right)
\stackrel{{\zeta}\uparrow0}{\longrightarrow}
\left(\begin{array}{c}
f'\;sf/(s+f)\\ 0 \\ s\end{array}\right)+O(k^2)
\end{equation}
and 
\begin{equation}
\label{3035}
\phi_1(0)=\frac{vk-s}{k}~.
\end{equation}

Now compare the mode $(\sigma_E,\rho_E,\psi_E)$ of infinitesimal 
change of far field $E_\infty$ from Eqs.\ (\ref{sigE}), 
(\ref{2033}) and (\ref{2034}) to the present perturbation mode 
in the limit of small $k$. After identifying
\be
\label{scale}
(\sigma_1,\rho_1,\psi_1)=(s\sigma_E,s\rho_E,s\psi_E),
\ee
the equations and boundary conditions for the modes are identical 
in leading order of the small parameter $s$. Therefore
the two modes have to become identical in the limit $s\ll f(v)<v$. 
Integration over $\psi_E$ 
yields for the electric potential $\phi_E(0)-\phi_E(-\infty)
=\int_{-\infty}^0dx\;\psi_E(x)$. This expression has to be 
of order unity since all other quantities are of order unity.
But this implies that $\phi_1(0)$ due to (\ref{scale}) has
to be of order $s$. Now compare the result for $\phi_1(0)$ 
in (\ref{3035}) which appears to depend in a singular way like
$1/k$ on the small parameter $k$. But for small $k$ and $s$ 
the expression $(vk-s)/k$ indeed can be of order $s$, namely if
\begin{equation}
\label{s=vk}
s=vk+O(k^3)~~~\mbox{for }~~k\ll\alpha(v)~.
\end{equation}
This fixes the dispersion relation $s=s(k)$ in the limit of small $k$.
The asymptote (\ref{s=vk}) is included as a solid line in Fig.~2.

\subsection{Physical interpretation of the small $k$ asymptote}

This result has an immediate physical interpretation: for small $k$,
the wave length of the transversal perturbation $2\pi/k$ 
is the largest length scale of the problem. It is much larger
than the thickness of the screening charge layer that is shown in Fig.~1. 
Therefore on the scale $1/k$, the charged front layer is very thin
and has the character of a surface charge rather than of a volume charge.
This surface is equipotential according to (\ref{sol+}) in linear 
approximation in the perturbation $\delta$, since
\be
&&\phi(x,\zeta=0,t)=\delta\;\frac{vk-s}{k}\;e^{ikx+st}+O(\delta^2)
\nonumber\\
&&\qquad\qquad=O(\delta k)+O(\delta^2),
\ee
if we insert the dispersion relation $s=vk$ from (\ref{s=vk}). 
The corresponding electric field ahead of the interface is 
\be
\label{E}
{\bf E}(x,\zeta=0^+,t)= -\big(v+\delta\; vk\;e^{ikx+st}\big)\;\hat\zeta
+O(\delta^2)
\ee
in the same approximation. The small $k$ limit of the ionization front
therefore is equivalent to an equipotential
interface at position $\zeta=0$, i.e., at a position
\be
\label{z}
z(x,t)=vt+\delta\;e^{ikx+st}
\ee
in the rest frame $z$ (\ref{zeta}). Its velocity in the $z$ direction
is therefore
\be
\label{v}
v(x,t)=\partial_t z(x,t)=v+\delta \;vk\;e^{ikx+st},
\ee
where $s=vk$ was inserted.
Comparison of (\ref{E}) and (\ref{v}) shows that the interface 
moves precisely with the electron drift velocity $v=-E$ within
the local field $E$.

We conclude that a linear perturbation of the ionization front 
whose wave length is much larger than all other lengths, 
has the same evolution as an equipotential interface ($\phi={\rm const.}$)
whose velocity is the local electron drift velocity $v=\nabla\phi$.
It exhibits the familiar Laplacian interfacial instability $s\propto k$.

\subsection{Asymptotics for large $k$ and arbitrary $E_\infty<0$}

For large wave-vector $k$, the numerical results for the dispersion
relation $s(k)$ in a field $E_\infty=-1$ approach a positive 
saturation value. We will now argue that the saturation
value is given by $s(k)=f(E_\infty)/2$. This asymptotic value,
which for $|E_\infty|=1$ equals $e^{-1}/2=0.184$, is included as a solid 
asymptotic line in Fig.\ 2.

When the electron and ion densities remain bounded, the equations with
the most rapid variation in (\ref{Matr1})--(\ref{Matr2}) for $k\gg1$
are given by
\begin{eqnarray}
\label{bigk}
\partial_{\zeta} \psi_1  &=& k^2 \phi_1+k^2\;E(\zeta)+O(k^0), \nonumber\\
\partial_{\zeta} \phi_1 &=& \psi_1
\end{eqnarray}
On the short length scale $2\pi/k$, the unperturbed electric field 
for $\zeta<0$ can be approximated as in (\ref{E-}) by
\begin{eqnarray}
E(\zeta) = -v -f(v)\zeta+O(\zeta^2),
\end{eqnarray}
so the equation for $\phi_1$ becomes
\be
\label{phi1}
\partial_\zeta^2\phi_1=k^2\Big(\phi_1-v-f(v)\zeta\Big).
\ee
The boundary condition (\ref{Init}) fixes $\phi_1(0)=(vk-s)/k$ and 
$\psi_1(0)=\partial_\zeta\phi_1=s$. The unique solution of (\ref{phi1})
with these initial conditions is
\begin{eqnarray}
\phi_1(\zeta) = v + f(v)\zeta - \frac{f(v)}{2k}\; e^{k\zeta} 
+ \frac{f(v)-2s}{2k}\; e^{-k\zeta}
\end{eqnarray}
for $\zeta<0$. Now the mode $e^{-k\zeta}$ would increase rapidly 
towards decreasing $\zeta$, create diverging electric fields 
in the ionized region and could not be balanced by any other terms 
in the equations. Therefore it has to be absent. The demand that its 
coefficient $(f(v)-2s)/2k$ vanishes, fixes the dispersion relation
\be
\label{s=f/2}
s(k)=\frac{f(v)}{2}+O\left(\frac1{k}\right)~~~\mbox{for }~~k\gg\alpha(v)~,
\ee
which convincingly fits the numerical results for large $k$ in Fig.~2.

\subsection{Physical interpretation of the large $k$ asymptote}

Also for this result
a physical interpretation can be given. First note that the 
$z$ component of the electric field on the discontinuity is
\be
\label{E2}
E_z(x,\zeta=0,t)= -\big(v+\delta\; s\;e^{ikx+st}+O(\delta^2)\big)
\ee
with $s=f(v)/2$. This is easily determined from either (\ref{sol+})
or (\ref{pertu}). Reasoning as in (\ref{E})--(\ref{v}),
we again find that the shock line of the electron density moves
with the local electron drift velocity --- as it should.

Second, one needs to understand why the electric field on the
shock line takes the particular form (\ref{E2}). In the frame 
$\xi=z-vt$ of the unperturbed front (\ref{zeta}), the electric field 
at the discontinuity is
\be
\label{E3}
E_z(x,\xi=\Delta,t)= -\left(v+\frac{f(v)}{2}\;\Delta+O(\Delta^2)\right),
\ee
where its position deviates with $\Delta(x,t)=\delta\;e^{ikx+st}$
from the planar front.

In linear perturbation theory, the amplitude $\delta$ of the perturbation
has to be much smaller than its wave length $2\pi/k$. Since this wave
length $2\pi/k$ now is much smaller than the width of the front, 
the linear perturbation $\Delta$ explores only a small region 
around the position of the shock front. In this region, the electric 
field of the unperturbed front is according to (\ref{E-}) approximated by
\be
\label{E4}
E_{z\:0}(\xi)=
\left\{\begin{array}{ll}-\big(v+f(v)\xi
+O(\xi^2)\big)~~&\mbox{for }~\xi<0\\
-v~~&\mbox{for }~\xi>0\end {array}\right. .
\ee
Therefore the electric field (\ref{E3}) is just the average 
over the behavior (\ref{E4}) for $\xi>0$ and $\xi<0$. 
This spatial averaging is enforced by the harmonic analysis
of linear perturbation theory that will suppress different growth
rates of positive or negative half-waves of the perturbation.

\subsection{A conjecture for the large $k$ asymptote}

We therefore conjecture: if the electric field of an unperturbed
front is 
\be
\label{Ec}
E_0(\xi)=
\left\{\begin{array}{ll}
-\big(v+a\xi+O(\xi^2)\big)~~&\mbox{for }~\xi<0\\
-\big(v+b\xi+O(\xi^2)\big)~~&\mbox{for }~\xi>0
\end{array}\right. 
\ee
near the position of the discontinuity $\xi=0$,
then a linear perturbation of this discontinuity with large $k$ 
will grow with rate
\be
s=\frac{a+b}{2}.
\ee 
If true, this behavior would have a stabilizing effect on large $k$ 
perturbations with growing curvature of the fronts, since the electric 
field decays in the non-ionized region ahead of a curved front, 
therefore $b<0$.


\section{Conclusions and outlook}

We have studied the (in)stability of planar negative ionization fronts
against linear perturbations. Such perturbations can be decomposed
into transversal Fourier modes. We have determined the dispersion
relation $s=s(k)$ shown in Fig.~2 numerically for a fixed field 
$E_\infty=-1$ far ahead of the front, and we have derived 
the analytical asymptotes
\begin{eqnarray}
\label{sk}
s&=&\left\{\begin{array}{ll}
|E_\infty| \;k &~~\mbox{ for }k\ll\alpha(|E_\infty|)/2\\
|E_\infty|\;\alpha(|E_\infty|)/2 &~~\mbox{ for }k\gg\alpha(|E_\infty|)/2
\end{array}\right. 
\end{eqnarray}
for arbitrary $E_\infty<0$.
Since we have studied the minimal model, there is only one
inherent length scale, namely the thickness of the charged layer
as shown in Fig.~1. This thickness is approximated by $1/\alpha(E_\infty)$.
The wave length $1/k$ of the Fourier perturbation therefore
has to be compared with this single intrinsic length scale 
$1/\alpha(E_\infty)$ of the problem.

A specific property of our calculation is the expansion about
a discontinuity of the electron density. Therefore we work 
in a coordinate system $\zeta=z-vt-\delta e^{ikx+st}$ (\ref{zeta}) 
that precisely follows
the position of the discontinuity, and we explicitly distinguish
in all calculations the non-ionized region $\zeta>0$ from the
ionized region $\zeta<0$. For the non-ionized region $\zeta>0$, 
there is an exact analytical solution (\ref{sol+}) for any $s$ and $k$
which determines the values of the fields at $\zeta=0$ as given
in (\ref{Init}). Eq.\ (\ref{Init}) serves as an initial condition
for the integration towards $\zeta<0$. The approach towards
$\zeta\to-\infty$ according to (\ref{End}) and (\ref{a4}) determines
the growth rate $s$ as a function of $k$. In general, this calculation 
has to be performed numerically with results as shown in Fig.~2. 
The limits of small and large $k$ can be derived analytically.
For small $k$, we can identify the perturbation mode with the mode
of infinitesimal change of $E_\infty$. For large $k$, the growth
rate corresponds to the evolution of the discontinuity in the 
unperturbed electric field averaged across the discontinuity. 
Both limits therefore have a simple physical interpretation.

The aim of the work was to identify a regularization for the 
interfacial model as suggested in \cite{ME,Firsov} 
and treated in \cite{Bernard}. Indeed, we have found that a Fourier 
mode for large $k$ in a far field $E_\infty=-v$ does not continue 
to increase with rate $s=vk$, but saturates at a value $s=f(v)/2$.
Still this is a positive value, and whether this suffices to
regularize the moving boundary problem, remains an open question.

Besides this one, future work will have to investigate two more 
questions. First of all, there is the ``simple'' possibility to extend
the model by diffusion. Diffusion is certainly going to 
suppress the growth rate of Fourier modes with large $k$
as our preliminary numerical work indicates.
But there is also a second more subtle and interesting possibility:
the growth rate of Fourier perturbations with large $k$ could
change for a curved front, as we have conjectured in Section IV.F. 
There we have argued that
the saturating growth rate $s=f(v)/2$ results from the average
over the slope $-f(v)$ of the field in the ionized region
and the slope 0 of the field in the non-ionized region.
For a curved front, the electric field in the non-ionized
region will have a slope of opposite sign that is proportional 
to the local curvature. We therefore expect the growth rate 
of a perturbation to decrease with growing 
curvature. These questions require future investigation.

\begin{acknowledgments}
M.A.\ gratefully acknowledges hospitality of CWI Amsterdam 
and a grant of the EU-TMR-network ``Patterns, Noise and Chaos''.
\end{acknowledgments}


\end{document}